\documentclass[shortAfour,times,sagev]{sagej}

\usepackage{physics}
\usepackage{dsfont}
\usepackage{array}
\usepackage{mathtools}
\usepackage{mathptmx}
\usepackage{hyperref}
\usepackage{tikz}
\usepackage{algorithm}
\usepackage{algpseudocode}
\usepackage{amsmath}
\usepackage{orcidlink}

\usetikzlibrary{quantikz2}

\def\journalname{SIMULATION}

\newcommand{\vacuum}{\ket{\mathbf{0}}}
\newcommand{\avacuum}{\bra{\mathbf{0}}}

\newenvironment{orcid}[1]%
{\subsection*{\normalsize\sagesf\bfseries ORCID iDs}\begin{refsize}\noindent #1}%
{\end{refsize}}

\begin{document}

\runninghead{D.~Emmanuel-Costa and M.~Epping}

\title{Hard-core Bosons in Action: Applications to Quantum Circuits}

\author{David Emmanuel-Costa\affilnum{1} and Michael Epping\affilnum{1}}
\affiliation{\affilnum{1}Institute of Software Technology, German Aerospace Center (DLR), Cologne, Germany}

\corrauth{David Emmanuel-Costa, Institute of Software Technology, German Aerospace Center (DLR), Linder Höhe, 51147 Cologne, Germany.}
\email{david.dacosta@dlr.de}

\begin{abstract}
The use of algebraic frameworks based on complex Clifford algebras for the representation and simulation of quantum circuits has been discussed in the literature. Recently, an alternative algebraic approach employing hard-core bosons has been proposed. Hard-core bosons provide a natural representation of multi-qubit systems, in which the tensor-product structure is realized directly and no sign corrections are required, in contrast to realizations based on complex Clifford algebras. Although both approaches are formally equivalent, the hard-core boson formulation exhibits computational advantages. This work reviews and extends the hard-core boson algebra for circuit simulation and presents an efficient implementation. A performance comparison with IBM Qiskit shows substantially improved execution times for simulations. Moreover, a new application is introduced in which the hard-core boson formalism is combined with genetic algorithms for quantum circuit synthesis.
\end{abstract}

\keywords{hard-core bosons, quantum computing, quantum simulators, quantum circuit synthesis, genetic algorithms}

\maketitle

\section{Introduction}
\label{sec:intro} 

Simulating quantum computers with conventional hardware is important for better understanding and developing quantum software~\cite{cicero2024simulationquantumcomputersreview}.
A key advantage is the ability to step through quantum circuits, inspect state vectors and probability distributions, and catch logic errors before deploying on noisy, real-world devices. In particular, state vector simulators are excellent for understanding and designing how multiple quantum algorithms fit together. 
Moreover, they provide a practical framework for error mitigation: researchers can model noise, prototype techniques (e.g., zero-noise extrapolation, readout correction), and validate workflows before moving to hardware. However, due to the complexity of the simulation, full state vector simulators typically scale only to $\approx35$ qubits on a workstation or 48 qubits on supercomputers~\cite{DeRaedt:2018btb}.

These simulation efforts are motivated by the rapid development of quantum computing, originally proposed by Feynman~\cite{Feynman:1981tf} (with earlier connections between quantum mechanics and computation in Refs.~\cite{benioff1980,Manin1980}) and later formalized by Deutsch~\cite{Deutsch:1989aa}. Quantum computing is a multidisciplinary field spanning computer science, physics, and mathematics and exploits quantum mechanics to solve certain problems faster than classical computers; in some cases, tasks can be executed exponentially faster on a quantum processor~\cite{Feynman:1981tf}, which is often referred as quantum supremacy~\cite{Arute2019}. Although the modern quantum computation has reached the noisy intermediate-scale quantum (NISQ) era~\cite{Preskill2018}, one is still at early stages for transiting to a fault-tolerant era, which is essential to enable quantum advantage. A comprehensive recent overview is given in Ref.~\cite{Basermann:2024oqq}. In this work, we focus on the circuit model of quantum computation, where quantum algorithms are described as sequences of gates (unitary operators) acting on wires (qubits), analogous to classical circuits~\cite{Nielsen_Chuang_2010}, and it has been shown that any unitary operation can be decomposed into single- and two-qubit gates~\cite{barenco1995elementary}.

An important aspect to consider when designing a state vector simulator is how to represent quantum circuits. Particularly noteworthy are representations based on algebraic expressions in complex Clifford algebras~\cite{Cafaro:2010wj,Hrdina:2022iti,Biswas:2024faj,Veyrac:2024jyo} and in hard-core boson algebras~\cite{Emmanuel-Costa:2025gog}. In the finite-dimensional complex Clifford-algebra approach, the central idea is to use the \emph{geometric product} as a unified primitive for representing qubits and gates~\cite{Hrdina:2022iti}. The geometric product of any two elements $u$ and $v$ of the algebra is defined by
\begin{equation}
  \label{eq:geometriprod}
  u\,v \;=\; u \cdot v \;+\; u \wedge v,
\end{equation}
where $\cdot$ denotes the (symmetric) scalar product and $\wedge$ the (antisymmetric) exterior product. In this algebra, the Witt-basis elements $f_i$ and $f_i^{\dagger}$ satisfy
\begin{equation}
  \label{eq:geometry}
  \qty{f_i^{\phantom{\dagger}},f_j^{\phantom{\dagger}}} \;=\; \qty{f_i^{\dagger},f_j^{\dagger}} \;=\; 0,
  \qquad
  \qty{f_i^{\phantom{\dagger}},f_j^{\dagger}} \;=\; \delta_{ij},
\end{equation}
and are used to define the state vectors needed to represent all qubits of a given quantum computer (see Ref.~\cite{Brackx:2008xnp} for details). Note that the only product in~\eqref{eq:geometry} is the geometric product; the elements $f_i$ and $f_i^{\dagger}$ thus behave like fermionic annihilation and creation operators.

For the simplest complex Clifford algebra with a single pair $f,f^{\dagger}$, the computational basis can be chosen as
\begin{equation*}
  \ket{0} \coloneqq f\,f^{\dagger},
  \qquad
  \ket{1} \coloneqq f^{\dagger}\ket{0} \;=\; f^{\dagger},
\end{equation*}
among other equivalent choices, and a qubit is any linear combination of the states $\ket{0}$ and $\ket{1}$. The inner product is defined by
$\braket{a}{b} \coloneqq \tfrac{1}{2}\,(a\,b)_0$, where $(\cdot)_0$ denotes projection onto the scalar (grade-zero) part of an element of the algebra. Any gate in this construction can be expressed in terms of $f_i$ and $f_i^{\dagger}$~\cite{Hrdina:2022iti}; for example, the NOT gate is
\begin{equation*}
  X \;=\; f \;+\; f^{\dagger}.
\end{equation*}
Remarkably, within this framework the geometric product encodes not only the composition of unitary operators but also the action of a gate on a state~\cite{Hrdina:2022iti}.

The generalization to multi-qubit states is straightforward. Define
\begin{equation*}
  \ket{00\cdots 0}
  \;\coloneqq\;
  \ket{0}_1 \otimes \ket{0}_2 \otimes \cdots \otimes \ket{0}_n
  \;=\;
  \prod_{i=1}^{n} f_i f_i^{\dagger}.
\end{equation*}
Thus, the tensor product is realized via the geometric product. In general, for any
$\lambda_i \in \{\,f_i,\; f_i^{\dagger},\; f_i f_i^{\dagger},\; f_i^{\dagger} f_i\,\}$,
one has the correspondence
\begin{equation}
  \label{eq:difficulty}
  \lambda_1 \otimes \cdots \otimes \lambda_n
  \;\longmapsto\;
  (-1)^{p}\,\lambda_1\cdots\lambda_n,
\end{equation}
where $p$ counts the sign changes induced by reordering graded elements; explicitly,
$p$ is the sum of the cardinalities of the sets
$S_i=\{\,j<i:\;\lambda_j\in\{f_j,\;f_j^{\dagger}f_j\}\,\}$
whenever $\lambda_i \in \{f_i,\;f_i^{\dagger}\}$. This sign arises because factors acting on different tensor components commute, whereas their geometric-product representatives anticommute. Hard-core boson algebras~\cite{Emmanuel-Costa:2025gog} provide an equivalent alternative that alleviates the sign bookkeeping in Equation~\eqref{eq:difficulty} while retaining the desirable properties of the finite-dimensional Clifford model.  In contrast to the complex Clifford-algebra construction~\cite{Hrdina:2022iti}, the hard-core boson algebra makes the composition of tensor-product operators naturally commutative, thereby simplifying the determination of the parity factor $(-1)^p$ in Equation~\eqref{eq:difficulty}.

Both algebraic approaches provide an elegant, unified representation of full circuits, and numerical implementations have been demonstrated~\cite{Hrdina:2022pho,Hrdina:2023gry,Emmanuel-Costa:2025gog}. In fact, the anticommutation relations in Equation~\eqref{eq:geometry} are powerful enough to carry out calculations without explicit matrices—the so-called \emph{oscillator expansion} technique~\cite{Wilczek:1981iz}. This feature underlies the central role of creation and annihilation operators in quantum mechanics~\cite{Dirac:1927dy} and many other areas~\cite{Mohapatra:1979nn,Wilczek:1981iz}; see, for example, Ref.~\cite{Cardoso:2015gfa} for algebraic simplifications using this technique.
\medskip

The present paper is an extended version of the prior work on the use of hard-core boson algebra~\cite{Emmanuel-Costa:2025gog} for simulating quantum computing.  After formalizing the hard-core boson framework, we develop the state vector simulator expressed within the hard-core boson algebraic approach. We also investigate the inverse problem, i.e., given a simplified hard-core boson algebraic expression, involving only
identity and creation operators, one determines, making use of genetic algorithms, a minimal-length quantum circuit over a chosen finite universal gate set that prepares the same final state.

The remainder of this paper is organized as follows. In the next section, one revisit the hard-core boson algebra in the context of qubits. The design of quantum circuits is then introduced. In the Section~\ref{sec:quantum-circuits} the representation of quantum gates based on products and sums of hard-core boson operators is established. Some applications are presented in Section~\ref{sec:applications}. A genetic algorithm for quantum circuit synthesis based on the hard-core boson algebra is developed in Section~\ref{sec:synthesis}. Finally, conclusions are drawn in Section~\ref{sec:conclusions}.

\section{Revisiting the Algebra of Hard-core Bosons}
\label{sec:hcb}

In this section the algebra of the hard-core bosons and its application to represent quantum circuits are revisited~\cite{Emmanuel-Costa:2025gog}. Hard-core bosons appeared motivated in many problems of condense matter physics and a few examples of application can be seen in~Refs.\cite{matsubara1956lattice,anderson1958absence,lieb1961two}. It will be demonstrated that hard-core bosons are a natural representation of qubit systems.

The state of a single qubit systems can be described by linear combinations of the orthonormal computational basis states $\{\ket0,\ket1\}$. Since the set $\{\ket0,\ket1\}$ is a basis, any linear operator acting on the state vectors can be fully decomposed in terms of four operators, namely,
$\ket{0}\bra{0}$, $\ket{1}\bra{1}$, $\ket{0}\bra{1}$ and $\ket{1}\bra{0}$.
The first two operators are the usual basis projectors, which sum to the identity,
\begin{equation*}
	\ket{0}\bra{0} \,+\, \ket{1}\bra{1} \,=\, \mathds{1},
\end{equation*}
while the remaining two operators are, in this context, identified with the annihilation and creation operators,
\begin{equation}
	\label{eq:opsdef}
	a\coloneqq\ket{0}\bra{1},\quad a^{\dagger}\coloneqq\ket{1}\bra{0},
\end{equation}
respectively.
Their names arise from the properties 
\begin{equation*}
	a \ket{0} = 0, \quad a \ket{1} = \ket{0}, \quad a^{\dagger} \ket{0} = \ket{1}, \quad a^{\dagger} \ket{1} = 0.
\end{equation*}
The projectors $\ket{0}\bra{0}$ and  $\ket{1}\bra{1}$ can be written as products of annihilation and creation operators, as
\begin{equation}
	\label{eq:projectors}
	aa^{\dagger}=\ket{0}\bra{0},\quad a^{\dagger}a=\ket{1}\bra{1}.
\end{equation}
The combination $N\coloneqq a^{\dagger}a$ is usually called the number operator and together with $M\coloneqq aa^{\dagger}$, it satisfies:
\begin{equation}
    \label{eq:NM}
    N+M=\mathds{1}\quad\text{and}\quad NM=MN=0,
\end{equation}
The matrix representations of the operators $N,M$ are idempotent Hermitian matrices with trace~$1$, hence each is a rank-one projector.

Taking into account Equations~\eqref{eq:opsdef} and~\eqref{eq:projectors}, one concludes that any linear operator $O$ can be described entirely in terms of annihilation and creation operators,
\begin{equation}
	\label{eq:op}
	O= O_{00}\, aa^{\dagger}  \,+\, O_{01}\, a  \,+\, O_{10}\, a^{\dagger}  \,+\, O_{11}\, a^{\dagger}a.
\end{equation}
with $O_{ij}\coloneqq\bra{i}O\ket{j}$.
Directly from  Equations~\eqref{eq:opsdef} and~\eqref{eq:projectors}, one writes a closed relation between the annihilation and  creation operators as
\begin{equation}
    \label{eq:subalgebra}
	\left\{ a,a^{\dagger}\right\}=\mathds{1} ,\quad a^2 \,=\, \left(a^{\dagger}\right)^2 \,=\,0.
\end{equation}
These operator properties form a noncommutative algebra, $\mathds{1}$ being the identity element. This algebra is usually called Grassmann algebra by physicists. One sees that any product of annihilation and creation operators always leads to one of the elements of the set  $\{1,\,0,\,a,\,a^{\dagger},\,aa^{\dagger},\,a^{\dagger}a\}$.

In a multi-qubit system of dimension $2^n$ one represents the basis states $\{\ket{b}\}$, with $b=0,\dots,2^n-1$ as
tensor product of the two state quantum system $\ket{0}$ and $\ket{1}$ as
\begin{equation*}
	\ket{b}\equiv\ket{b_1\,b_2\,\cdots\,b_n} = \ket{b_1}\otimes\ket{b_2}\otimes\cdots\otimes\ket{b_n},
\end{equation*}
where the elements $b_i$ are the bits in the binary basis of the integer $b$ (the bit on the far right $b_n$ is the least significant bit), and therefore $b_i\in\{0,1\}$. 
In the context of a multi-qubit vector space, it is natural to extend the definition of the annihilation and creation operators by the following definitions
\begin{align*}
		a_i           & \coloneqq \mathds{1} \otimes \cdots \otimes \mathds{1} \otimes a^{\phantom{\dagger}} \otimes \mathds{1} \otimes \cdots\otimes\mathds{1}, \\
		a^{\dagger}_i & \coloneqq \mathds{1} \otimes \cdots \otimes \mathds{1} \otimes a^{\dagger} \otimes \mathds{1}  \otimes\cdots\otimes\mathds{1},
\end{align*}
where in the above equations each annihilation or creation is set at the position $i$ in the tensor product string,
while the other position are set to the identity. One can now deduce the algebra for the extended annihilation, $a_i$,  and creation, $a^{\dagger}_i$,  operators.  Setting $i\neq j$, one has
\begin{equation}
	\label{eq:hcb}
	\left[a^{\phantom{\dagger}}_i, a_j\right]=\left[a^{\dagger}_i, a^{\dagger}_j\right]=\left[a_i, a^{\dagger}_j\right]=0.
\end{equation}
These relations above are typical of bosonic commutation relations. It means, that operators that act on different qubit indices commute, i.e. the commutativity of the tensor product of operators is automatically preserved. This is a neat difference when comparing with the Clifford algebra construction given in Equation~\eqref{eq:geometry} in the introduction. The relation of annihilation and creation operators on the same qubit index, i.e.,  $i=j$, is instead given by  anti-commutation relations,
\begin{equation}
	\label{eq:hcb:anti}
	\left\{ a_i,a_i^{\dagger}\right\}=\mathds{1},\quad
	\left\{ a_i^{\phantom{\dagger}},a_i\right\}=\left\{ a_i^{\dagger},a_i^{\dagger}\right\}=0,
\end{equation}
exhibiting now a fermion-like behavior. This mixed behavior, or this 
inconsistency of the qubits with the properties of bosons and fermions, is what is called hard-core bosons in the literature and their algebra is fully determined by Equations~\eqref{eq:hcb} and~\eqref{eq:hcb:anti}. In qubit systems, hard-core bosons are locally the same as parafermions~\cite{Bonatsos:1999wv,Wu:2001zro}. Since each operator $a^{\phantom{\dagger}}_i$ and $a^{\dagger}_i$ acts on the $i$-th qubit subspace, the relations given in Equations~\eqref{eq:hcb} and ~\eqref{eq:hcb:anti} can be regarded as defining an abstract algebra, which can be interpreted either as an operator product or as a tensor product. 

The full basis of the multi-qubit space can now be systematically enumerated. In what follows, one denotes the vacuum state vector $\vacuum \coloneqq \ket{00\cdots0}$. The transformation of the annihilation and creation operators when acting on the computational basis vectors is straightforward (in expressions like $\ket{b_1\cdots x\cdots b_n}$ it is assumed that $x$ is at the position $i$):
\begin{align*}
		a_i\ket{b_1\cdots0\cdots b_n} & =0,\\
		a_i\ket{b_1\cdots1\cdots b_n} &= \ket{b_1\cdots0\cdots b_n},\\
		a^{\dagger}_i\ket{b_1\cdots0\cdots b_n} &=\ket{b_1\cdots1\cdots b_n}, \\
		a^{\dagger}_i\ket{b_1\cdots1\cdots b_n} & =0.
\end{align*}
Therefore any state vector $\ket{b_1b_2\cdots b_n}$ is written as
\begin{equation}
	\label{eq:ket}
	\ket{b_1b_2\cdots b_n} = \Big(a^{\dagger}_1\Big)^{b_1} \Big(a^{\dagger}_2\Big)^{b_2} \cdots \Big(a_n^{\dagger}\Big)^{b_n} \vacuum.
\end{equation}
Here $b_i$ again stands for the value of the bit at position~$i$. One can verify that such state vectors are orthogonal
and properly normalized, i.e.,
\begin{equation*}
	\bra{b_1' b'_2\cdots b'_n}\ket{b_1b_2\cdots b_n} = \delta_{b_1b'_1}\delta_{b_2b'_2}\cdots\delta_{b_nb'_n}.
\end{equation*}
For computations, it is useful to rewrite Equation~\eqref{eq:ket} without introducing powers as 
\begin{equation}
	\label{eq:rel1}
	\begin{aligned}
	\ket{b_1b_2\cdots b_n} 
	=& \qty[b_1\,a^{\dagger}_1+(1-b_1)\mathds{1}]
	\Big[b_2\,a^{\dagger}_2+(1-b_2)\mathds{1}\Big] \times \\ &\cdots\times\Big[b_n\,a^{\dagger}_n+(1-b_n)\mathds{1}\Big]\vacuum.
	\end{aligned}
\end{equation}

This derivation demonstrates that a hard-core boson structure naturally emerges from a multi-qubit space, with the advantageous property of preserving tensor product commutativity. Since there is no need to correct the tensor parity, as required in Equation~\eqref{eq:geometry}, this implementation appears more efficient for constructing a quantum computer simulator. However, for a large number of qubits, the exponential growth of the multi-qubit space presents significant challenges for simulation. It is worth to point out that this is just a reformulation of standard matrix representations that will not magically reduce the complexity of simulations, but it seems to promise some practical advantages when it comes to implementations.

Since between different indices all operators commute, the product of an arbitrary number of operators can be performed by associating operators with same indices together, maintaining their relative position. As an example, one has
\begin{equation*}
	a_5\,a_2^{\dagger}a_1\, a_2\, a_3^{\dagger}\,a_2^{\dagger} \,=\,
	a_1\,a_2^{\dagger}a_2a_2^{\dagger}\,a_3^{\dagger}\,a_5 \,=\,
	a_1\,a_2\,a_3^{\dagger}\,a_5.
\end{equation*}
An ordering is very useful to compare two different expressions.
We first order by qubit index and then  the creation operator before the annihilation operator on the same qubit, using the anti-commutation relation (\ref{eq:hcb:anti}).
For example,
\begin{equation*}
	a_5\,a_1\, a_2\, a_3^{\dagger}\,a_2^{\dagger} \,=\, a_1\,a_3^{\dagger}\,a_5 \,-\, a_1\,a_2^{\dagger}a_2\,a_3^{\dagger}\,a_5.
\end{equation*}
Similarly, the Hermitian conjugation in the  context of hard-core bosons is relatively straightforward, one needs to interchange individual $a_i\leftrightarrow a^{\dagger}_i$ while the combinations $a_i a^{\dagger}_i$ and $a^{\dagger}_i a_i$ remain unchanged. An example of Hermitian conjugation is
\begin{equation*}
	\left(a_1\,a_2^{\dagger}a_2\,a_3^{\dagger}\,a_5\right)^{\dagger}
	\,=\, a_1^{\dagger}\,a_2^{\dagger}a_2\,a_3\,a_5^{\dagger}.
\end{equation*}
Thanks to the commutativity of operators acting on different qubits, it is now straightforward to simplify algebraic expressions. 
In contrast to this, operators acting on different qubits in the Clifford-algebra representation anticommute. 
Removing the need to keep track of these resulting minus signs potentially leads to faster computations in the hard-core boson representation. A demonstration of the equivalence of this hard-core boson algebra to the Clifford algebra~\cite{Hrdina:2022iti} is given in Appendix~\ref{sec:equivalence}. The equivalence of this approach to the Pauli-string methods~\cite{Nielsen_Chuang_2010,Biswas:2024faj} is then given in Appendix~\ref{sec:pauli}.
\medskip

Before closing this section, it is worth noting that the projectors $N,\,M$ defined in Equation~\eqref{eq:NM} induce single–qubit projectors
\begin{equation*}
N_i \coloneqq a_i^{\dagger}a_i^{\phantom{\dagger}} 
\qquad\text{and}\qquad
M_i \coloneqq a_i^{\phantom{\dagger}}a_i^{\dagger},
\end{equation*}
acting on qubit $i$ and trivially on all other qubits. The operators $N_i,\,M_i$ are projectors on the $i$-th copy of $\mathbb{C}^2$ in the tensor product $(\mathbb{C}^2)^{\otimes n}$ and they commute for all $i$ and $j$. When $i=j$, they satisfy 
\begin{equation*}
    N_i+M_i=\mathds{1} 
    \qquad\text{and}\qquad
    N_i M_i = M_i N_i = 0 .
\end{equation*}
Using the above relations, one obtains a complete resolution of the identity on the $2^n$-dimensional multi-qubit space
\begin{equation*}
    \mathds{1}
    = \prod_{i=0}^{n-1}\bigl(M_i+N_i\bigr)
    = \sum_{b=0}^{2^{n}-1}\mathds{P}_b ,
\end{equation*}
where the projectors $\mathds{P}_b$ are defined by
\begin{equation*}
    \mathds{P}_b \coloneqq \prod_{i=0}^{n-1} N_i^{b_i}\,M_i^{1-b_i},
\end{equation*}
with $b_i\in\{0,1\}$ denoting the $i$-th bit in the binary expansion of the integer $b$, where $0\leq b<2^n$. By construction, $\{\mathds{P}_b\}$ is a complete set of mutually orthogonal projectors:
\begin{equation*}
    \mathds{P}_b^2=\mathds{P}_b,
    \qquad
    \mathds{P}_b\,\mathds{P}_c=\delta_{bc}\,\mathds{P}_b,
    \qquad
    \sum_{b=0}^{2^{n}-1}\mathds{P}_b=\mathds{1}.
\end{equation*}
Since there are $2^n$ such projectors, any product of the commuting operators $N_i$ and $M_i$ can be written as a linear combination of the projectors $\mathds{P}_b$. Hence $\operatorname{span}\{\mathds{P}_b\}$ is the subspace of all commutative elements~\cite{budinich:2019}. 

\section{Quantum Circuit Simulation}
\label{sec:quantum-circuits}

\sbox{0}{\begin{quantikz}
		& \gate{X}  & \qw
	\end{quantikz}}
\sbox{1}{\begin{quantikz}
		& \gate{Y}  & \qw
	\end{quantikz}}
\sbox{2}{\begin{quantikz}
		& \gate{Z}  & \qw
	\end{quantikz}}
\sbox{3}{\begin{quantikz}
		& \gate{H}  & \qw
	\end{quantikz}}
\sbox{4}{\begin{quantikz}
		& \gate{S}  & \qw
	\end{quantikz}}
\sbox{5}{\begin{quantikz}
		& \gate{T}  & \qw
	\end{quantikz}}
\sbox{6}{\begin{quantikz}
		& \gate{R_x(\theta)}  & \qw
	\end{quantikz}}
\sbox{7}{\begin{quantikz}
		& \gate{R_y(\theta)}  & \qw
	\end{quantikz}}
\sbox{8}{\begin{quantikz}
		& \gate{R_z(\theta)}  & \qw
	\end{quantikz}}

\begin{table}
  \small\sf\centering
  \caption{\label{tab:gates} The most common single-qubit quantum gates written in terms of creation and annihilation operators. Note that qubit indices were dropped from the table for better readability.}
  \begin{tabular}{cc}
    \toprule
    \textbf{Quantum Circuits} & \textbf{Operator Sequences} \\
    \midrule
    \usebox{0}                & $a^{\dagger}+a$                             \\
		\usebox{1}                & $i(a^{\dagger}-a)$                                                                               \\
		\usebox{2}                & $aa^{\dagger}-a^{\dagger}a$                                                                      \\
		\usebox{3}                & $\frac1{\sqrt{2}}\qty(aa^{\dagger}-a^{\dagger}a+a^{\dagger}+a)$                                   \\
		\usebox{4}                & $aa^{\dagger}+ia^{\dagger}a$                                                                     \\
		\usebox{5}                & $aa^{\dagger}+e^{i\tfrac{\pi}4}a^{\dagger}a$                                                     \\
		\usebox{6}                & $\cos(\tfrac{\theta}2)\qty(aa^{\dagger}+a^{\dagger}a)-i\sin(\tfrac{\theta}2)\qty(a+a^{\dagger})$ \\
		\usebox{7}                & $\cos(\tfrac{\theta}2)\qty(aa^{\dagger}+a^{\dagger}a)+\sin(\tfrac{\theta}2)\qty(a^{\dagger}-a)$  \\
		\usebox{8}                & $e^{-i\tfrac{\theta}2}aa^{\dagger}+e^{i\tfrac{\theta}2}a^{\dagger}a$                             \\
\bottomrule
\end{tabular}
\end{table}

Any 1-qubit gate can readily be expressed from Equation~\eqref{eq:op} in terms of the operators $a$, $a^{\dagger}$, $N$ and $M$. Also any multi-qubit gate can be expressed in terms of the operators $a_i^{\phantom{\dagger}}$, $a_i^{\dagger}$, $N_i$ and $M_i$, by extending Equation~\eqref{eq:op}. 
In the $2^n$-dimensional multi-qubit space, a unitary operator acting on the space can be interpreted as
\begin{equation*}
	U\,=\,\sum_{b,b^{\prime}=0}^{2^n-1}U_{bb^{\prime}}\,\ket{b^{\,}}\bra{b^{\prime}},
\end{equation*}
where each operator $\ket{b}\bra{b^{\prime}}$ can be written as
\begin{equation*}
	\ket{b^{\phantom{\prime}}\!}\bra{b^{\prime}} \,=\,\ket{b^{\phantom{\prime}}_1\cdots b^{\,}_n}\bra{b^{\prime}_1\cdots b^{\prime}_n}=
	\ket{b^{\,}_1}\bra{b^{\prime}_1}\otimes\cdots\otimes\ket{b^{\,}_n}\bra{b^{\prime}_n}.
\end{equation*}
Then, for each $\ket{b_1}\bra{b^{\prime}_1}$ one identifies the correct operator expression by making use of Equations~\eqref{eq:opsdef} and~\eqref{eq:projectors}. As illustration, a general control gate,
\begin{equation*}
\Lambda^1(U)=\ket{0}\bra{0}_c\otimes\mathds{1}+\ket{1}\bra{1}_c\otimes U,
\end{equation*}
between the first (control) and second qubit (target), in terms of annihilation and creation operators reads 
\begin{equation}
\label{eq:control_U}
	\Lambda^1(U)=M_c\,+\,N_c\,U,
\end{equation}
where $c$ stands for the control qubit and $U$ is a unitary operator acting on $\mathbb{C}^2$, the target subspace. Explicating the above equation in terms of creation and annihilation operators yields 
\begin{equation*}
	\Lambda^1(U)=a_{c}a_c^{\dagger}+a_c^{\dagger}a_c\qty(U_{11}a_{t}a_t^{\dagger}+U_{12}a_t+U_{21}a_t^{\dagger}+U_{22}a_t^{\dagger}a_t).
\end{equation*}
With $U=\exp(i\,H)$, where $H$ is an Hermitian operator (Hamiltonian), one obtains
\begin{equation*}
	\Lambda^1(U) =M_c\,+\,N_c\left(1+\sum_{n=1}^{\infty}\frac{(i\,H_t)^n}{n!}\right)=
	 e^{i\,N_c\otimes\, H_t}.
\end{equation*}
It is worth to point out that Equation~\eqref{eq:control_U} can be extended to a multi-controlled-U gate as
\begin{equation*}
	\Lambda^m(U)=\prod_{i=1}^{m}\left(M_i + N_i\right)\,+\,\left(\prod_{i=1}^m N_{i}\right)\otimes \Bigl(U_t - \mathds{1}_t\Bigr),
\end{equation*}
where it was assumed that the first $m$ qubits are control qubits and $t>m$ is the target qubit\footnote{Note that the index $t$ in $U_t$ or $ \mathds{1}_t$ enforces that $U$ or $ \mathds{1}$ acts only on the target qubit, respectively.}. Finally, Table~\ref{tab:2qubits} shows examples of two-qubit gates written only in terms of annihilation and creation operators.

In order to understand the circuit synthesis below, let us briefly review the concept of coherence~\cite{Streltsov2017}. Generally speaking coherence is the ability to interfere, i.e. the ability to form superpositions of states, which may lead to constructive or destructive interference. Whether a state is considered in superposition or not is basis dependent. It is a convention to call the elements of the computational basis incoherent and by extension any diagonal density matrix. All other states are superpositions of them, i.e. contain coherence. 
The coherence of a state can be quantified using a mathematical measure. This can be, for example, the sum of the absolute values of all off-diagonal elements of the density matrix~\cite{Baumgratz2014}.
Operations that map incoherent states to incoherent states are called incoherent. In the resource theory of coherence, these operations are considered 'free'. As these operations do not make use of interference, the most important feature of quantum computers, they can be directly implemented on classical computers (as stochastic processes). Indeed shifting these computations from a quantum computer to the classical computer at state preparation or measurement can be formalized, as was done by one of the authors in Ref.~\cite{Epping:2022agh}. This allows simplifying the quantum circuit and the synthesis can prefer circuits that allow for this.

\sbox{0}{\begin{quantikz}[column sep=10pt, row sep={20pt,between origins}]
		& \swap{1} & \qw \\
		& \targX{} & \qw
	\end{quantikz}}
\sbox{1}{\begin{quantikz}[column sep=10pt, row sep={20pt,between origins}]
		& \ctrl{1} & \qw \\
		& \targ{} &  \qw
	\end{quantikz}}
\sbox{2}{\begin{quantikz}[column sep=10pt, row sep={20pt,between origins}]
		& \ctrl{1} &  \qw \\
		&  \ctrl{0}&  \qw
	\end{quantikz}}
\sbox{3}{\begin{quantikz}[column sep=10pt, row sep={20pt,between origins}]
       & \ctrl{2} & \qw \\
       & \ctrl{1} & \qw\\
       & \targ{} & \qw
    \end{quantikz}}	

\begin{table}
  \small\sf\centering
  \caption{\label{tab:2qubits} Some two or more qubit quantum gates written with creation and annihilation operators.}
	\begin{tabular}{cc}
		\toprule
		\textbf{Quantum Circuits} & \textbf{Operator Sequences}                                                                           \\
		\midrule
		\usebox{0}                & $a_1^{\phantom{\dagger}}a_1^{\dagger}a_2^{\phantom{\dagger}}a_2^{\dagger}+a_1^{\dagger}a_1^{\phantom{\dagger}}a_2^{\dagger}a_2^{\phantom{\dagger}}+a_1^{\dagger}a_2^{\phantom{\dagger}}+a_1^{\phantom{\dagger}}a_2^{\dagger}$ \\
		\usebox{1}                & $a_1^{\phantom{\dagger}}a_1^{\dagger}+a_1^{\dagger}a_1^{\phantom{\dagger}}\qty(a_2^{\dagger}+a_2^{\phantom{\dagger}})$ \\
		\usebox{2}                & $a_1^{\phantom{\dagger}}a_1^{\dagger}+a_1^{\dagger}a_1^{\phantom{\dagger}}\qty(a^{\phantom{\dagger}}_2a_2^{\dagger}-a_2^{\dagger}a_2^{\phantom{\dagger}})$ \\
		\usebox{3}               &
	    $a_1^{\phantom{\dagger}}a_1^{\dagger}+
	    a_1^{\dagger}a_1^{\phantom{\dagger}}a_2^{\phantom{\dagger}}a_2^{\dagger}+
        a_1^{\dagger}a_1^{\phantom{\dagger}}a_2^{\dagger}a_2^{\phantom{\dagger}}\qty(a_3^{\dagger}+a_3^{\phantom{\dagger}})$ \\
		\bottomrule
	\end{tabular}
\end{table}

We now demonstrate the power of the \emph{oscillator expansion} technique,i.e., the algebraic relations between  annihilation and creation operators, to simulate probabilities for a given circuit. As is standard in quantum computers, the initial state of the computation is set to the all-zero state, which we call the vacuum state $\vacuum$ in the context of hard-core bosons. 
This is not a restriction, since any initial state can be prepared and included at the beginning of the gate sequence. To make sense of the described hard-core boson formalism for simulating a given quantum circuit, some assumptions must be made. 
For a given sequence of gates $G\coloneqq G_{p}G_{p-1}\cdots G_1$, 
one assumes these gates to operate sequentially on the initial state $\vacuum$, such that the initial state evolves into a final state $\ket{f}$. 
It is also assumed that during the application of the gates $G_i$, no measurements are performed. 
The measurement only occurs after the full sequence of gates, marking the end of the computation. 
Finally, for the sake of simplicity, it is assumed that the quantum simulation operates under ideal conditions, without considering any source of errors. 

Within this algebraic framework, the effect of the full circuit $G$ is calculated by using the operations allowed by the hard-core boson algebra. The strength of this method is to calculate $G$ without requiring the determination of any matrix elements for the operators involved, relying solely on the rules given in Equation~\eqref{eq:hcb} and~\eqref{eq:hcb:anti}.

One of the most common practices of quantum simulations is the determination of the final state vector just before a final measurement. However, in a real quantum computing device, measurements are the only means to extract information about the final state. The circuit must be executed repeatedly to obtain the probability distribution of the final state. Theoretically, the probability to observe the state $\ket{b}$ after applying $G$ to $\vacuum$ is given by the Born rule
\begin{equation*}
	P(b\mid G)\coloneqq \left|\bra{b}G\vacuum\right|^2.
\end{equation*}
Using the expression given in Equation~\eqref{eq:rel1} in the bracket of the above equation, the probability $P(b,G)$ becomes the expectation value of an algebraic expression (defined in the hard-core boson algebra) with respect to $\vacuum$, as
\begin{equation}
\label{eq:master}
P(b \mid G)
= \Bigl\lvert
 \avacuum
 \prod_{k=1}^{n} \qty[\,b_k\,a_k + (1-b_k)\,\mathds{1}\,]\,
 G \,\vacuum
 \Bigr\rvert^{2}.
\end{equation}
The algebraic rules given in Equations~\eqref{eq:hcb} and~\eqref{eq:hcb:anti} allow to move the creation operators to the left of the expression, and the annihilation operators to the right. Any term with an annihilation operator vanishes when applied to the state $\vacuum$ and any creation operator also vanishes when applied to the left to the dual state $\avacuum$. The only terms that persist are terms proportional to the identity, due to the Kronecker tensor in the anticommutator relations given in Equation~\eqref{eq:hcb:anti}. This demonstrates the potential of this algebraic technique for simulating quantum circuits.

It is worth pointing out that one can simplify the scheme above with the aim to extract the probabilities without systematically preparing the dual state $\bra{b}$. Before applying $G$ to the state $\vacuum$, one should order the expression by moving all annihilation operators to the right. As previously, all terms containing annihilation operators vanish when applied to $\vacuum$, resulting in the final state
\begin{equation}
	\label{eq:special}
	\ket{f}\,\equiv\, G\vacuum \,=\,\left( \sum^{2^n-1}_{b=0} \zeta_b\,(a_1^{\dagger})^{b_1}(a_2^{\dagger})^{b_2}\cdots (a^{\dagger}_n)^{b_n}\right)\vacuum,
\end{equation}
where $\zeta_b$ are just complex coefficients. For each $b$ and its binary decomposition $b=(b_1,b_2\dots b_n)$, one can read the probability directly as
\begin{equation*}
	P(b|G)= |\zeta_b|^2.
\end{equation*}
The operator inside of the parenthesis on the right-hand side of Equation~(\ref{eq:special}) is no longer unitary, but it allows to reconstruct the probabilities. The resulting state is then given by
\begin{equation*}
	\ket{f}= \sum^{2^n-1}_{b=0} \zeta_b\ket{b}.
\end{equation*}

One can use the method described above to compare two distinct quantum circuits. Let us take two circuits, with respective final unitaries $G$ and $G^{\prime}$, acting on vacuum $\vacuum$, the two circuits are equivalent, up to an irrelevant global phase, if and only if
\begin{equation*}
	\left|\avacuum G^{\dagger}\, G'\vacuum\right|^2=1.
\end{equation*}
Similar to what was done in Equation~\eqref{eq:master}, the ordering of creation operators to the left and annihilation operators to the right ensures that all terms vanish, except those proportional to the identity. This simplifies the evaluation of the bracket, making it calculable within the hard-core boson formalism.

\section{Applications}
\label{sec:applications}

In this section, we apply our formalism to some standard quantum algorithms to illustrate the potential of circuit representations and simulations through the hard-core boson algebra, which was the subject of the previous section. The three selected classes of quantum algorithms are: preparation of the Greenberger–Horne–Zeilinger (GHZ) state, Grover’s algorithm, and the quantum Fourier transform (QFT). We follow the algorithm definitions and conventions adopted by the Munich Quantum Toolkit Benchmark Library (MQTBench)~\cite{quetschlich2023mqtbench}. In order to test the viability of oscillator expansion techniques based on hard-core bosons for representing larger quantum circuits, a library written in \verb!C++! called Quipo is being developed~\cite{quipo}. Performance tests of the Quipo library are carried out using the state vector simulator of the Quantum Information Software Kit (Qiskit)~\cite{qiskit2024}, version 2.0.0, as the reference.

\subsection{The Greenberger–Horne–Zeilinger state}

\begin{figure}
	\centering
	\begin{quantikz}
		\lstick{\(\ket{0}\)} & \gate{H} & \ctrl{1} & \ctrl{2} & \qw \\
		\lstick{\(\ket{0}\)} & \qw & \targ{} & \qw & \qw \\
		\lstick{\(\ket{0}\)} & \qw & \qw & \targ{} & \qw
	\end{quantikz}
	\caption{A quantum circuit that prepares the Greenberger–Horne–Zeilinger.}
	\label{tab:GHZ}
\end{figure}
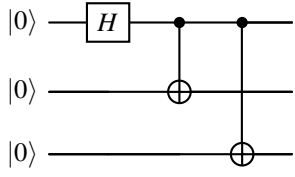

\begin{figure}
	\centering
	\framebox{\includegraphics[width=0.9\linewidth]{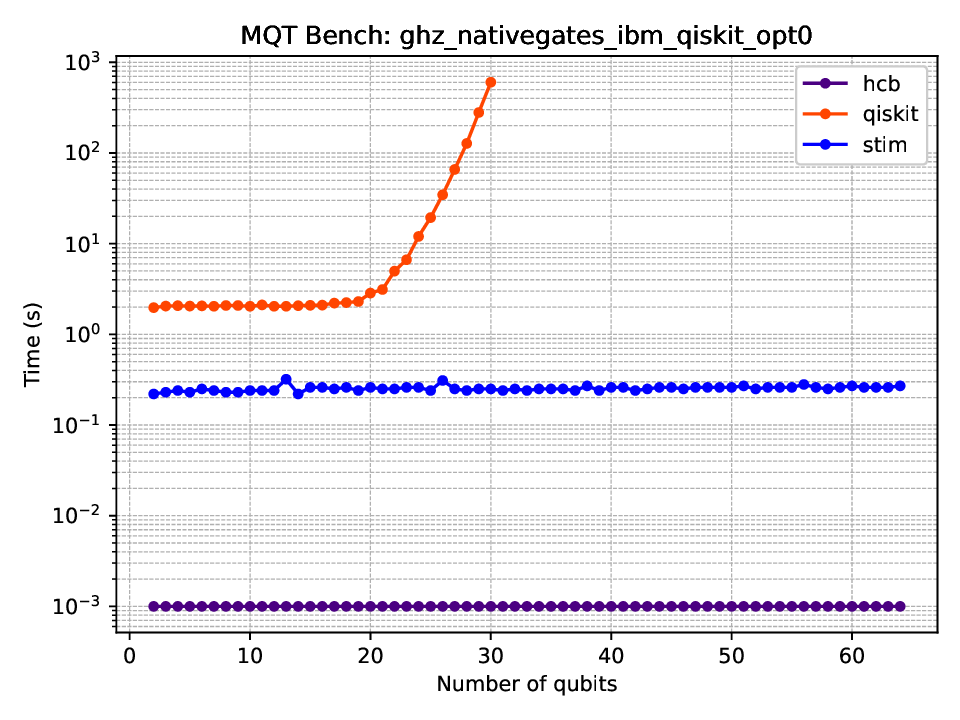}}
	\caption{The Greenberger-Horne-Zeilinger state for different qubit number provided by the Munich Quantum Toolkit Benchmark Library, up to 64 qubits. The figure presents execution times for three simulators were used, namely, Quipo~\cite{quipo} (hard-core boson implementation), Qiskit~\cite{ qiskit2024} and Stim~\cite{gidney2021stim}.}
	\label{fig:GHZ}
\end{figure}

In this example, we first consider a 3-qubit circuit that prepares the GHZ state. The GHZ state is a certain type of entangled quantum state that involves at least three~\cite{merminQuantumMysteriesRevisited1990} or more qubits~\cite{Greenberger:1989tfe}. 
The 3-qubit GHZ state has been realized experimentally~\cite{Bouwmeester:1998iz}. It plays a crucial role in fundamental tests of quantum mechanics versus local realism and in many quantum information and quantum communication schemes~\cite{Epping:2017imv}. Figure~\ref{tab:GHZ} shows the quantum circuit that prepares the GHZ state. Using Equation~\eqref{eq:special} for the circuit given in Figure~\ref{tab:GHZ}, on gets
\begin{equation*}
	G_{\text{GHZ}}\vacuum \,=\, \left(\frac{1}{\sqrt{2}}\mathds{1}\,+\,\frac{1}{\sqrt{2}}a_1^{\dagger}a_2^{\dagger}a_3^{\dagger}\right)\vacuum.
\end{equation*}
As expected from Section~\ref{sec:quantum-circuits}, the right-hand side of the above equation is no longer a unitary operator, but leads to the same final vector state,
\begin{equation*}
	\ket{\text{GHZ}}=\frac{\ket{000}+\ket{111}}{\sqrt{2}}.
\end{equation*}
One reads from the GHZ state all nonvanishing probabilities, which are $P(000|G_{\text{GHZ}})=P(111|G_{\text{GHZ}})=50\%.$ In our numerics, we extended this to higher number of qubits, i.e.
\begin{equation*}
	\ket{\text{GHZ}}=\frac{\ket{00\cdots0}+\ket{11\cdots1}}{\sqrt{2}}.
\end{equation*}
For this goal, we have used quantum circuits from the Munich Quantum Toolkit Benchmark Library~\cite{quetschlich2023mqtbench}, which are given as OpenQASM files~\cite{Cross:2017bqf}, version 2.0, up to 64 qubits. We took the state vector simulator provided by Qiskit~\cite{qiskit2024} as the reference. Then we used our library Quipo~\cite{quipo} that implements our simulator, which is designed using the hard-core boson algebra. Figure~\ref{fig:GHZ} shows the execution times for simulating the circuits that prepare the GHZ state, running on laptops with processor Core Ultra 7 and Apple M1 (2020). We conclude that Quipo's computation times are at least three order of magnitude smaller than Qiskit's.

Note that the GHZ is prepared using only Clifford gates, which are known to be perfectly simulated in polynomial time on a classical computer, as a result of the Gottesman–Knill theorem~\cite{gottesman1998,Aaronson:2004xuh}. The Qiskit curve shows how the performance decreases heavily and the simulation became infeasible for more than 30 qubits on a laptop. For this reason, in our numerical analysis we have used the tool Stim~\cite{gidney2021stim}, which can  efficiently simulate Clifford circuits. We can see in Figure~\ref{fig:GHZ}, Quipo also performs better than the Stim simulator. 

\subsection{Grover's algorithm}

\begin{figure}[tbp]
  \centering
  \begin{quantikz}[column sep = 0.5cm]
  \ket{+} & \gate[2]{U_\omega} & \gate[2]{U_D} & \meter{}\\
  \ket{+} & & & \meter{}
  \end{quantikz}
  \caption{
     Quantum circuit for the Grover algorithm to find the solution $\ket{\omega}=\ket{10}$ marked by the oracle $U_\omega$, followed by the action of the diffusor $U_D$ (the reflection across the initial state). Note that since $n=2$ qubits are used, the Grover iteration $U_D U_\omega$ is applied only once.}
  \label{fig:grover}
\end{figure}
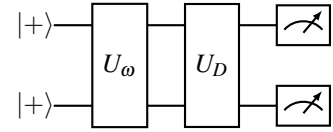

In this example, the hard-core boson algebraic formalism is applied to simulate an instance of Grover's algorithm~\cite{Grover:1996rk}. The goal of Grover's algorithm is to find the solution state $\ket{\omega}\in \mathds{C}_2^{\otimes n}$ which is marked by the oracle $U_\omega$ via a phase flip, i.e., $U_\omega \ket{\omega} = - \ket{\omega}$. To illustrate the hard core boson formalism, Grover's algorithm is simulated on $n=2$ qubits for $\ket{\omega}=\ket{10}$. The circuit is depicted in Figure~\ref{fig:grover} and the involved operations are
\begin{equation*}
   U_\omega = \mathds{1}-2\proj{\omega} = 1 - 2 a_1^\dagger a_1 a_2 a_2^\dagger
\end{equation*}
and
\begin{equation*}
U_D = 2\proj{+}-\mathds{1} = \frac{1}{4}(1 + a_1 + a_1^\dagger)(1 + a_2 + a_2^\dagger)-1.
\end{equation*}
The action of the circuit is
\begin{equation*}
 \begin{aligned}
  C\,\coloneqq\, & U_D U_\omega H^{\otimes 2} \\
  = & a_1 a_2^\dagger a_2 \,+\, a_1^\dagger \,-\, a_1^\dagger a_2^\dagger a_2 + a_1^\dagger a_1 a_2 \,+\, a_1^\dagger a_1 a_2^\dagger \,-\, a_2,
 \end{aligned}
\end{equation*}
where the preparation of the Hadamard state $\ket{+}\coloneqq(\ket{0}+\ket{1})/\sqrt{2}$ was also included. From the procedure to deduce the final state in Equation~\eqref{eq:special} in the Section~\ref{sec:quantum-circuits}, one gets $C\vacuum= a_1^\dagger\vacuum$, which leads to the outcome probability 
\begin{equation*}
  P(2 | C) = |\avacuum a_1C \ket{0}|^2 = |\avacuum a_1 a_1^\dagger\vacuum|^2 = 1.
\end{equation*}
As expected, the solution $\ket{\omega}=\ket{10}$ is correctly obtained with a success probability of~$100\%$.

\subsection{Quantum Fourier transform algorithm}

\begin{figure}
	\centering
	\framebox{\includegraphics[width=0.9\linewidth]{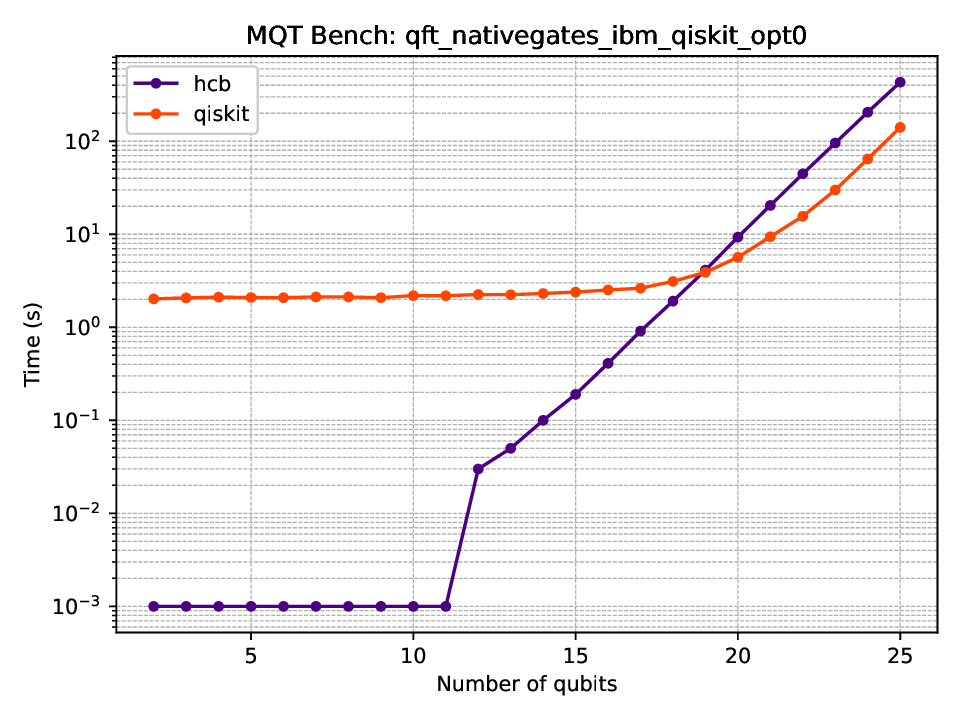}}
	\caption{Quantum Fourier transformation for different qubit number provided by the Munich Quantum Toolkit Benchmark Library, up to 25 qubits. The figure presents execution times for three simulators were used, namely, Quipo~\cite{quipo} (hard-core boson implementation) and  Qiskit~\cite{ qiskit2024}.}
	\label{fig:qft}
\end{figure}

The quantum Fourier transform (QFT) is the quantum analogue of the classical Fourier transform and has been playing an important role in the conception and development of many quantum algorithms. A classical Fourier transform maps a function to another function that describes the frequency content of the original function. The mathematical structure of the Fourier transform allows quantum algorithms to exploit quantum superposition and interference, yielding exponential speedups for certain problems and highlighting quantum computing’s advantage over classical methods.

\begin{figure}
	\centering
	\framebox{\includegraphics[width=0.9\linewidth]{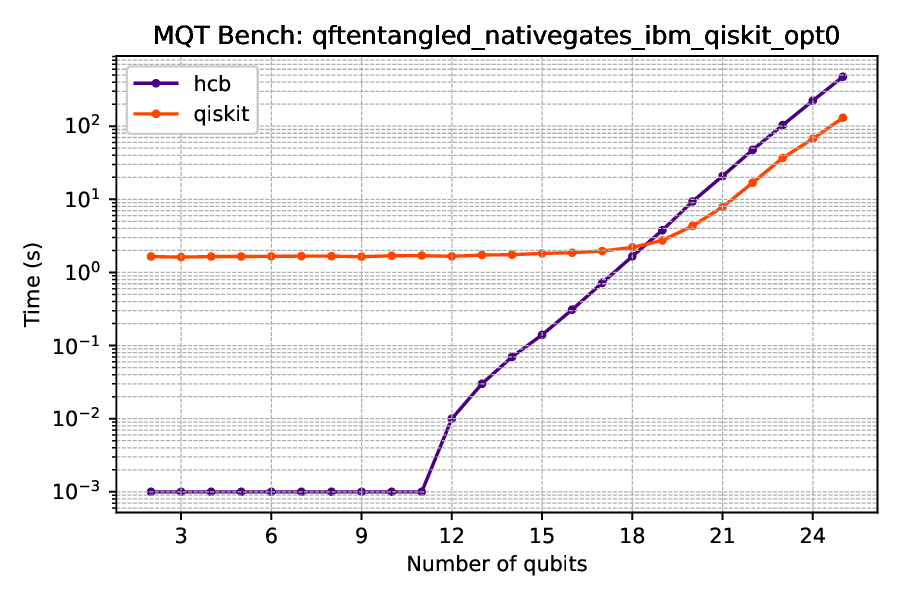}}
	\caption{Quantum Fourier transformation to entangled qubits for different qubit number provided by the Munich Quantum Toolkit Benchmark Library, up to 25 qubits. The figure presents execution times for three simulators were used, namely, Quipo~\cite{quipo} (hard-core boson implementation) and  Qiskit~\cite{ qiskit2024}.}
	\label{fig:qftentangled}
\end{figure}

In our numerics, we have also considered `entangled QFT quantum circuits', which apply regular QFT to entangled qubits. Both the QFT and the entangled QFT algorithms were taken from the Munich Quantum Toolkit Benchmark Library~\cite{quetschlich2023mqtbench}. For both types, we have selected from the MQTBench interface circuits for the target-dependent native IBM gate set $\{rz$, $sx$, $x$, $cx\}$ using up to 25~qubits, which delivered us the circuits as OpenQASM~2.0 files~\cite{Cross:2017bqf}. Both Quipo and  Qiskit tools were run on an Apple M1 (2020) with $16GB$ of RAM and under macOS Sequoia 15.3.2 (\verb!C++! compiler: Apple clang version 16.0.0). As an example, for 17 qubits the QFT algorithm performed $0.91s$ on Quipo and $1.22s$ on Qiskit, while the entangled QFT performed $0.82s$ on Quipo and $1.94s$ on Qiskit. We have verified that all the probabilities of the final state for each algorithm are exactly the same for both Quipo and Qiskit. 

In Figure~\eqref{fig:qft} we plot the execution time for the QFT circuits. The comparison is complicated by the fact that, in contrast to Qiskit, the Quipo library is a simple implementation of the hard-core algebra without exploring performance engineering and exploiting multiple CPUs. We have also performed a similar analysis for the entangled QFT, which is shown in Figure~\ref{fig:qftentangled}. One sees that the hard-core algebraic approach performs better than Qiskit up to 18 qubits. It was observed that the advantage of the Quipo over Qiskit depends strongly on the circuit topology. 

\section{Quantum Circuit Synthesis and Circuit Optimization}
\label{sec:synthesis}

The hard-core boson algebra expresses a given quantum circuit only using creation and annihilation operators and further simplifies it using only identity and creation operators as stated in Equation~\eqref{eq:special}. A natural question is whether this identification can be inverted: starting from a simplified hard-core boson expression, can we derive a minimal-length sequence of gates whose composition prepares the same final state vector? In general, this inverse mapping is not unique. A practical objective is therefore to perform circuit synthesis, or circuit extraction, with respect to a chosen universal gate set, so that an appropriate composition reproduces the target algebraic expression with as few gates as possible. In this context, we restrict ourselves to a finite gate set. Indeed, such a procedure could be useful for rewriting an initial circuit in terms of the universal gates available on a specific quantum device. This is usually what a compiler is used for.

A natural question is how to determine a minimal-length sequence of universal gates, and genetic algorithms (GA) provide a robust optimization framework~\cite{Holland1975,Goldberg1989} for such problems. GAs have been used in the context of quantum computing, e.g., a GA was used for state preparation on quantum computers~\cite{Creevey:2023qux}. We first give an intuitive description of GAs in biological terms and link them to circuit synthesis below. GAs are population-based, derivative-free optimizers that mimic natural selection: candidate solutions (chromosomes) are iteratively improved by selection, crossover, and mutation. Fitness-proportional reproduction favors high-quality candidates while maintaining diversity for global exploration, yielding robust, efficient minimization even on noisy, nonconvex landscapes. Moreover, the chromosome length refers to the number of genes within a chromosome that represent specific manufacturing campaigns in a genetic algorithm.

This is the approach we shall adopt. We start by fixing the universal gate set as $\{T,\, H,\, CX\}$. For convenience one may also include the gates $S=T^2$ and $X=H\,T^4\,H$ as primitives, i.e., $\{T,\, H,\, CX,\, S,\, X\}$. If one restricts to $n$-qubits and assuming all-to-all connectivity, there is a total of $n^2+3n$ different genes.

Within our construction, a chromosome encodes an ordered list of primitive gates (one gene per gate application). One interprets a chromosome as a sequence of quantum gates applied to the vacuum state producing a candidate output state vector to be scored by the GA’s fitness function. For example, the length-3 chromosome $H(1)-CX(1,2)-CX(1,3)$ corresponds to the state vector
\begin{equation*}
   CX(1,3)\,CX(1,2)\,H(1)\,\vacuum=\,
  \frac{\mathds{1}\,+\,a_1^{\dagger}a_2^{\dagger}a_3^{\dagger}}{\sqrt{2}}\vacuum.
\end{equation*}
While the length of chromosomes can vary dynamically to adapt to the capacity planning problem being addressed in the GA, we fix length within all GA execution for simplicity. In order to determine the shortest chromosome that achieves the target state, we wrap the GA in a binary search over chromosome length. It should be remark that it is not guaranteed to find a solution of that length even if it exists, the method strongly depends on the choice of parameters describe below (population size, mutation rate, etc.).

We now describe our adapted GA procedure in what follows. We identify the initial quantum circuit, $C_0$, given by a string of unitary operators, and implement a GA to search for a chromosome of a given length which is the best candidate for reproducing the state $\ket{C_0}\coloneqq C_0\vacuum$. We then use binary search to identify the minimal chromosome length that attains the target as close as possible within the tolerance~$\epsilon$. The GA method initiates by preparing randomly $N$ chromosomes of length $L$ - the initial population $P$. Next, one calculates the fitness of all elements in the initial population. 
The fitness function $f_t$ is defined for each chromosome $p\in P$ as
\begin{equation}
  \label{eq:fitf}
   f_t(p) = \left|\avacuum O_p^{\dagger}\, C_0\vacuum\right|,
\end{equation}
where $C_0$ is target circuit and $O_p$ the corresponding operators for the chromosome $p$. It is worth to remark that both quantities $C_0$ and $O_p$ are expressed in terms of creation operators and the identity. 
By definition, one has
\begin{equation*}
    0\leq f_t(p)\leq1.
\end{equation*}
The stop condition that we use in our GA implementation is given by 
\begin{equation*}
  \max_{ p\in P} f_t(p) \geq 1-\epsilon. 
\end{equation*}
The parameter $\epsilon$ is chosen to control the tolerance of the solution. Once the condition is satisfied, the best chromosome solution is the element of $P$ that maximizes the fitness function $f_t$. 

The choice of the fitness function $f_t$ plays an important role for the success of the GA. The function given in Equation~\eqref{eq:fitf} seems to be the best guess in this context, but it requires that any viable approximated chromosome, $p$ must correspond to $f_t(p)\simeq1$. Since we are interested in a proof of concept of using GA for quantum circuit synthesis, a detailed investigation for the best fitness function is beyond the scope of this paper.

Once the choice of the fitness function is fixed, the GA is describe by repeating a set of steps, which defines a generation, until the stop condition is satisfied. Within each generation, one considers the sequence of three operations, namely selection, crossover and mutation:
\begin{enumerate}
\item\textbf{Selection} \par The selection consists in discarding the elements of the current population $P$ with a fitness $f_t(p)$ below a given threshold. The threshold depends strongly on the choice of fitness function. In the case of the fitness function given in Equation~\eqref{eq:fitf}, the overall factor $1/\sqrt{2}$ of the Hadamard gate $H$, can easily value near $0.7$, yet it may still correspond to an inadequate chromosome.
\item[]
\item\textbf{Crossover} \par In sexual reproduction, as observed in the real world, the genetic material of two parents is mixed when their gametes merge. Usually, chromosomes are randomly split and merged, with the consequence that some genes of a child come from one parent while others come from the other parent—the so-called \emph{crossover mechanism}. In order to describe how crossover is realized in our GA, each chromosome, i.e., a string of gates, is written as 
\begin{equation*}
\mathbf{G}\to G_1-G2-\dots-G_{L-1}-G_L,  
\end{equation*}
where $L$ is the fixed chromosome length and a crossover probability $p_C\in(0,1]$ is also set. Thus, given two parents $\mathbf{A}$ and $\mathbf{B}$,
\begin{equation*}
\mathbf{A}\to A_1-\dots-A_L,\quad\text{and} \quad \mathbf{B}\to B_1-\dots-B_L,
\end{equation*}
 with probability $p_C$ we perform crossover; otherwise $1-p_C$ we pass the parents' chromosomes unchanged. For a  single-point crossover, we draw a cut \(k\in\{1,\dots,L-1\}\) uniformly and form offspring $\mathbf{o}^{1}$ and $\mathbf{o}^{2}$ such that
 \begin{equation*}
     \begin{aligned}
      \mathbf{o}^{1}&\to A_1-\dots-A_k-B_{k+1}-\dots-B_L,\\
      \mathbf{o}^{2}&\to B_1-\dots-B_k-A_{k+1}-\dots-A_L.
     \end{aligned}
 \end{equation*}

\item[]
\item\textbf{Mutation} \par The last ingredient of such GAs is mutation — the random deformation of the genetic information of an individual by means of
radioactive radiation or other environmental influences. In real reproduction, the probability that a certain gene is mutated is almost equal for all
genes. Therefore the following mutation technique is natural. With probability $p_M$ a single gene at a specific position $k$ is modified from $G_k$ to $G^{\prime}_k$, i.e. the string $G_1-\dots-G_k-\dots-G_L$ becomes $G_1-\dots-G^{\prime}_k-\dots-G_L$.
\end{enumerate}
 The best chromosome solution is returned when the GA terminates. Repeated gates, other than the $T$-gate, indicate that the length $L$ can be shortened. In the Algorithm~\ref{algo:GA} we sketch the adopted GA for circuit extraction.

\begin{algorithm}
\caption{\label{algo:GA} Genetic Algorithm for circuit extraction}
\begin{algorithmic}[1]
\Require population size $N$, crossover $p_C$, mutation $p_M$, fitness $f$, stop condition $\mathcal{C}(P)$
\State $P \gets \textsc{Init}(N)$; \ \textsc{Eval}$(P,f)$
\While{\textbf{not} $\mathcal{C}(P)$}
 \State $M \gets \textsc{Select}(P,N)$
 \State $O \gets \textsc{Crossover}(M,p_C)$
 \State \textsc{Mutate}$(O,p_M)$
 \State \textsc{Eval}$(O,f)$
 \State $P \gets \textsc{Replace}(O)$
\EndWhile
\State \Return \textsc{Best}$(P)$
\end{algorithmic}
\end{algorithm}

\begin{figure}
	\centering
	\framebox{\includegraphics[width=0.9\linewidth]{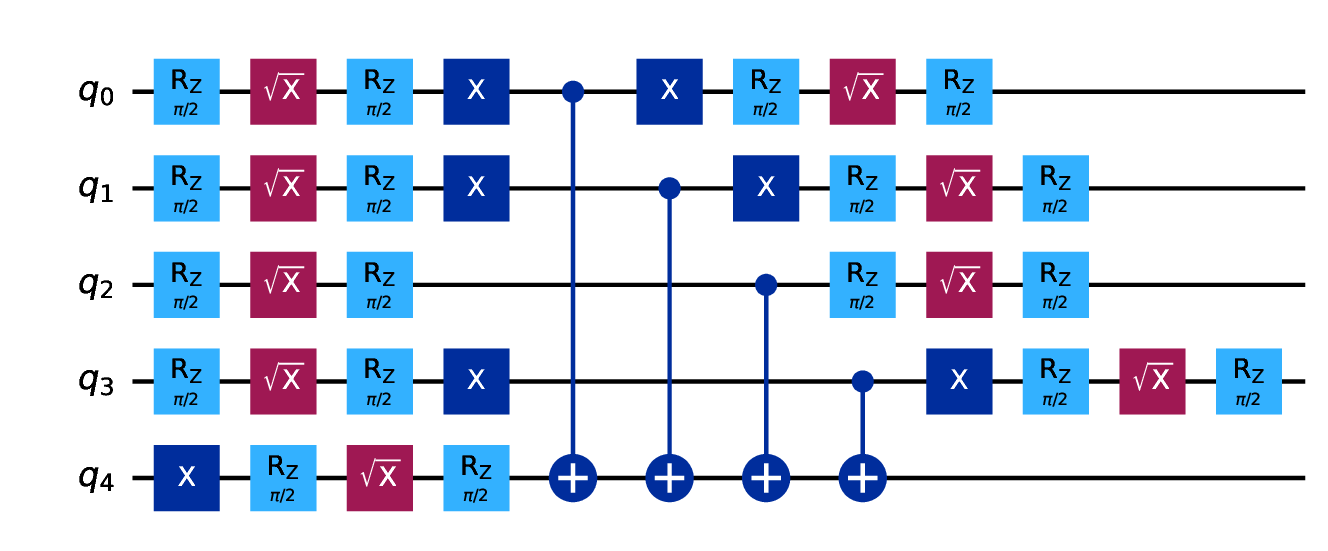}}
	\caption{Deutsch–Jozsa circuit for five input qubits and written in terms of primitive gates $\{RZ$, $SX$, $X$, $CX\}$. This quantum circuit was drawn using Qiskit~\cite{qiskit2024}.}
	\label{fig:deutsch}
\end{figure}

As proof of concept we apply this method to the Deutsch-Jozsa algorithm for 5~qubits. An OpenQASM for such circuit was taken from the Munich Quantum Toolkit Benchmark Library which for this example is expressed in terms of $\{RZ$, $SX$, $X$, $CX\}$. The corresponding quantum circuit is composed of 38 gates. We have implemented our GA within the \verb!C++! Quipo library, in particular for calculating the fitness function given in Equation~\eqref{eq:fitf}.  Applying our GA together with bisection method we get the following 6-gate sequence:
\begin{equation*}\small
    G\to X(1)-X(4)-X(3)-CX(4,0)-CX(3,2)-H(4).
\end{equation*}
In our numerics, we have selected the initial population size to be $10^6$ and a mutation rate $p_M=0.1$. We have not found any solution for the  chromosome length less than 6. The gate sequence above lead to the same statistic as the circuit given in Figure~\ref{fig:deutsch}, i.e.,
\begin{equation*}
     \left|\bra{\psi}G\vacuum\right|^2 \,=\,
    \left|\bra{\psi}C_0\vacuum\right|^2 \,=\, \frac{1}{2},
\end{equation*}
for $\ket{\psi}\in\{\ket{11111}$, $\ket{01111}\}$.

\section{Conclusions}
\label{sec:conclusions}

Quantum computing is an interdisciplinary field that mixes ideas from computer science, physics, and mathematics, and aims to simulate quantum systems more efficiently than with classical methods. In the usual model, we describe computations with quantum circuits built from qubits and quantum gates, giving researchers a common way to talk about quantum algorithms. These algorithms use uniquely quantum effects like superposition and entanglement, which are hard to mimic with classical computers. Notable exceptions are some special kinds of circuits, such as Clifford circuits, which can still be simulated efficiently on classical hardware. However, for more general circuits, one often relies on expensive state vector simulators to study their behavior and benchmark quantum algorithms on classical machines.

The performance of quantum simulations depends crucially on the representation of basic quantum information. Algebraic constructions based on complex Clifford algebras~\cite{Hrdina:2022iti} describe quantum circuits using only algebraic elements and the geometric product, but the non-Abelian nature of the geometric product requires parity corrections to recover the commutative tensor product. An alternative yet equivalent formulation uses the algebra of hard-core bosons~\cite{Emmanuel-Costa:2025gog} to refine this representation, realizing the tensor product without parity corrections because the product is commutative for elements with different indices.

The present paper reviews and extends the algebraic hard-core boson framework. Multi-qubit systems naturally fit into this algebra, making it well suited for quantum simulations. We gave the controlled-multiqubit gate description in the hard-core algebra, thereby completing the instruction set needed to run simulations based on this framework. It was also shown how to simulate a quantum computing process and compute the final state and it's measurement probabilities within this approach, using the \emph{oscillator expansion} technique through the annihilation and creation operator algebra. This approach recreates the tensor product as expected, and thus enables more efficient computational tools, such as the Quipo \verb!C++!~library~\cite{quipo}, for quantum system simulations. Taken together, the three examples considered in this work—namely, the preparation of GHZ states, Grover’s search algorithm, and the quantum Fourier transform—show that the hard-core boson algebra can faithfully reproduce standard quantum algorithms within a unified circuit framework. A comprehensive benchmarking of Quipo \verb!C++!~library~\cite{quipo} remains an important direction for future work.

A natural direction, taken in Section~\ref{sec:synthesis}, was to investigate the inverse problem: given a simplified hard-core boson expression (involving only identity and creation operators), construct a minimal-length quantum circuit over a chosen finite universal gate set that prepares the same final state, effectively using the algebraic hard-core boson framework as a circuit-synthesis and compilation tool. It was then shown that genetic algorithms can be used to implement the inverse problem. The procedure is still at an early stage, and further investigation is needed in order to fully compile quantum circuits to operator strings involving only the native gate set of a quantum device.

\appendix
\section{Equivalence to Complex Clifford Algebras}
\label{sec:equivalence}

One addresses now the question whether it is possible to map the hard-core boson algebra~\cite{Emmanuel-Costa:2025gog} to the fermionic-like  formulation in the context of complex Clifford algebras~\cite{Hrdina:2022iti}. In fact, there is a well known mapping between hard-core bosons and fermions, the so-called Jordan-Wigner
transformation~\cite{Jordan:1928wi}, in which the annihilation and creation operators are transformed as
\begin{equation*}
	a_i \to f_i =  \mathcal{J}_i a_i \quad\text{and}\quad
	a^{\dagger}_i \to f^{\dagger}_i = \mathcal{J}^{\dagger}_i a^{\dagger}_i,
\end{equation*}
where $\mathcal{J}_i$ is an unitary operator defined by
\begin{equation*}
	\mathcal{J}_i \coloneqq \prod^{i-1}_{k=1} e^{i\,\pi\,N_k}.
\end{equation*}
From the definition, one deduces the following relation
\begin{equation*}
	\mathcal{J}^{\dagger}_i = \mathcal{J}_i =\prod^{i-1}_{k=1}\left( -Z_k\right),
\end{equation*}
and if one takes into account that $Z_i$ anticommutes with $a_i$ and $a^{\dagger}_i$, it can then demonstrated
that the resulted operators $f_i$ and $f_i^{\dagger}$ fulfill the fermionic algebra given in~\eqref{eq:geometry},
i.e.,
\begin{equation*}
	\qty{f^{\phantom{\dagger}}_i,\,f^{\dagger}_j} \, =\, \delta_{ij}\mathds{1},
	\quad
	\qty{f^{\phantom{\dagger}}_i,\,f^{\phantom{\dagger}}_j} \, = \,\qty{f^{\dagger}_i,f^{\dagger}_j} \,=\, 0.
\end{equation*}
The Jordan-Wigner transformations are know to be an elegant and systematic way to include fermionic properties into quantum computations~\cite{Veyrac:2024jyo}.

\section{Pauli Strings}
\label{sec:pauli}

In this appendix, one establishes first the map between the annihilation and creation operators and the $X$- and $Y$-Pauli matrices.
\begin{equation}
    \label{eq:pauli:def}
	X_i = a_i\,+\,a_i^{\dagger},\quad
	Y_i = i\,(a_i^{\dagger}\,-\,a_i),
\end{equation}
with the converse being given by
\begin{equation*}
	a_i^{\phantom{\dagger}}  =\frac{X_i\,+\,i\,Y_i}{2},\quad
	a_i^{\dagger}  = \frac{X_i\,-\,i\,Y_i}{2}.
\end{equation*}
The Z-Pauli matrices are given by  $Z_i \coloneqq i\,X_i\,Y_i$ and one has
\begin{equation*}
Z_i = a_i^{\dagger}a_i\,-\,a_ia_i^{\dagger} \,=\,M_i\,-\,N_i = 1 - 2\, N_i = 2\, M_i -1.
\end{equation*}
There is a straightforward relation between $Z_i$ and the number operator $N_i$ and the operator $M_i$, as
\begin{equation*}
N_i=\frac{\mathds{1}-Z_i}{2},\,\quad  M_i=\frac{\mathds{1}+Z_i}{2}.
\end{equation*}
As an example, for any 2-dimensional linear operator, $O$, the expression given in~\eqref{eq:op} can be rewritten as linear combination of Pauli matrices $X,Y,Z$, and the identity matrix $\mathds{1}$, as 
\begin{multline*}
    	O=\frac{1}{2}\left[\,\left( O_{00} + O_{11} \right)\,\mathds{1} 
    	\,+\, \left( O_{10} + O_{01} \right)\,X \right.\\
		\left.
		\,+\, i\left( O_{10} - O_{01} \right)\,Y  
		\,+\, \left( O_{00} - O_{11} \right)\,Z\,\right].
\end{multline*}
Since the products of annihilation and creation operators are nothing more than a tensor product, it is not surprising that the definitions given in~\eqref{eq:pauli:def} for $i\neq j$ lead to 
\begin{align*}
    \left[ X_i,\,X_j\right]=&\left[ Y_i,\,Y_j\right]=\left[ Z_i,\,Z_j\right] \\
    =&\left[ X_i,\,Y_j\right]=\left[ X_i,\,Z_j\right]=\left[ X_i,\,Z_j\right]=0,
\end{align*}
which leads to the formalism of Pauli strings used extensively in quantum computation~\cite{Nielsen_Chuang_2010}. Pauli strings allows for a structured simulation process using classical computations~\cite{Biswas:2024faj}.

\begin{acks}
The author~D.E.C. would like to thank Dr.~Thomas Keitzl for valuable discussions and for carefully reading the manuscript.
\end{acks}

\begin{funding}
This project was made possible by the DLR Quantum Computing Initiative and the Federal Ministry of Research, Technology and Space; \href{https://qci.dlr.de/alqu/}{qci.dlr.de/projects/ALQU}.
\end{funding}

\begin{orcid}
David Emmanuel-Costa~\orcidlink{0000-0002-5068-7125} https://orcid.org/0000-0002-5068-7125\\
Michael Epping~\orcidlink{0000-0003-0950-6801} https://orcid.org/0000-0003-0950-6801
\end{orcid}

\bibliographystyle{SageV} 
\bibliography{references}

\begin{biogs}
\noindent\textbf{David Emmanuel-Costa} is a researcher at the German Aerospace Centre (DLR). He obtained his Ph.D.~degree in theoretical particle physics under the supervision of Gustavo Branco at the University of Lisbon (UL), Portugal. He worked as a postdoc at Deutsches Elektronen-Synchrotron (DESY) in Hamburg and at UL. His research interests include quantum computing optimization and simulation, and applications of quantum computing to particle physics (e.g., neutrino physics). His email address is \url{david.dacosta@dlr.de}.
\medskip

\noindent\textbf{Michael Epping} leads a research group at the German Aerospace Centre (DLR) that focuses on compilation and error correction for quantum computing. His main interest lies in getting the most out of DLR's quantum computer prototypes. He studied quantum information theory in Siegen, Vienna and in Düsseldorf, where he completed his Ph.D. under the supervision of Dagmar Bruß at the Heinrich-Heine University. You can contact him at~\url{michael.epping@dlr.de}.
\end{biogs}

\end{document}